\documentclass{article}
\usepackage[utf8]{inputenc}
\usepackage[T1]{fontenc}
\usepackage{booktabs}
\usepackage{graphicx}
\usepackage{authblk}
\usepackage{quoting}
\usepackage{framed}
\usepackage{multicol}
\usepackage[numbers]{natbib}
\usepackage{comment, url}
\usepackage{hyperref}
\hypersetup{
colorlinks=true,
linkcolor=blue,
filecolor=magenta,  
}
\title{SwissCovid: a critical analysis of\\ risk assessment by Swiss authorities}
\author[1]{Paul-Olivier Dehaye}
\author[2]{Joel Reardon}
\affil[1]{PersonalData.IO\thanks{The first author wishes to thank many PersonalData.IO and MyData volunteers, as well as multiple contributors to the GitHub repositories of the DP-3T collaboration \cite{DP3T}.}}
\affil[1]{MyData Geneva}
\affil[2]{University of Calgary}

\newenvironment{packed_item}{
\begin{itemize}
  \setlength{\itemsep}{1pt}
  \setlength{\parskip}{0pt}
  \setlength{\parsep}{0pt}
}{\end{itemize}}

\date{\today}
\begin{document}
\maketitle

\begin{abstract}
Ahead of the rollout of the SwissCovid contact tracing app, an official public
security test was performed. During this audit, Prof.~Serge Vaudenay and Dr.~Martin
Vuagnoux described a large set of problems with the app, including a new
variation of a known false-positive attack, leveraging a cryptographic weakness
in the Google and Apple Exposure Notification framework to tamper with the
emitted Bluetooth beacons. Separately, the first author described a
re-identification attack leveraging rogue apps or SDKs.

The response from the Swiss cybersecurity agency and the Swiss public health
authority was to claim these various attacks were unlikely as they required
physical proximity of the attacker with the target (although it was admitted the
attacker could be further than two meters). The physical presence of the
attacker in Switzerland was deemed significant as it would imply such attackers
would fall under the Swiss Criminal Code.

We show through one example that a much larger variety of adversaries must be
considered in the scenarios originally described and that these attacks can be
done by adversaries without any physical presence in Switzerland. This goes directly against official findings of Swiss public authorities evaluating the risks associated with SwissCovid. To move the discussion further along, we briefly discuss the growth of the attack surface and harms with COVID-19 and SwissCovid prevalence in the population.

While the focus of this article is on Switzerland, we emphasize the core technical findings and cybersecurity concerns are of relevance to many contact tracing efforts. 
\end{abstract}
\section{Context}
The SwissCovid app is the Swiss digital contact tracing app\footnote{On Android:
	\href{https://play.google.com/store/apps/details?id=ch.admin.bag.dp3t}{https://play.google.com/store/apps/details?id=ch.admin.bag.dp3t}}
for the COVID-19 crisis. It implements the Google and Apple Exposure
Notification (GAEN) APIs \cite{Apple, Google}. The protocol underlying those APIs was partly inspired by work done by a team called DP-3T, primarily based in Switzerland \cite{DP3T}. 

In May 2020, the SwissCovid app was deployed for testing to a select group of military
personnel and civil servants. In parallel, a public security
test\footnote{\href{https://www.melani.admin.ch/melani/en/home/public-security-test/scope_and_rules.html}{https://www.melani.admin.ch/melani/en/home/public-security-test/scope\_and\_rules.html}}
was initiated by the Reporting and Analysis Centre for Information Assurance
MELANI, the core of the Swiss National Cyber Security Centre (NCSC). The test itself was
hindered by the following factors:
\begin{itemize}
\item The reliance of the app on the Google and Apple API, because it made it
much less transparent for a complete security audit.
\item The (fortunate) lack of keys reported through the system to the central server, in order to indicate pseudonymously that an individual participant in the test had received a positive diagnostic to COVID-19.
\end{itemize}

On Saturday June 5th 2020, Prof.~Serge Vaudenay and Dr.~Martin Vuagnoux submitted a
report to MELANI~\cite{Vaudenay} detailing a long list of vulnerabilities\footnote{See \url{https://lasec.epfl.ch/people/vaudenay/swisscovid.html} for additional information by Vaudenay on the fate of his report.}. While many of those vulnerabilities had been described before, the report was helpful in documenting them impartially, without pulling into the discussion alternative protocols as strawmen. This report however also included a new vulnerability\footnote{We note that the Federal Office of Public Health asserts in \cite{FOPH} ``DP-3T researchers at EPFL and ETH Z\"{u}rich have evaluated this aspect of the June 5th report [by Vaudenay and Vuagnoux]. The researchers acknowledge this new variant was not previously evaluated.''}, relying on metadata tampering to affect the emission signal strength. This tampering of a few bits in the payload would have to be done blind as it would target data encrypted using AES-CTR\footnote{AES-CTR , or \emph{Advanced Encryption Standard-Counter mode}, is a standard for encrypting data. This standard does not authenticate data, which means data can be modified.}. 

Having read the report on June 6th 2020, the first author realized a larger set of actors that he had originally anticipated could actually leverage these weaknesses. He described his views in two letters to MELANI, on June 9th and June 11th\footnote{For completeness, this is referenced as [INR-4554] in the NCSC's official findings \cite{findings}.}.

On June 15th, the NCSC published their own detailed assessment \cite{NCSC} of those vulnerabilities. This was immediately followed by an assessment from the Swiss public health authority  FOPH \cite{FOPH}. 

Below we will describe the attacks, describe the assessments made by the Swiss authorities of the associated risks, and finally describe a potential attack vector that would directly invalidate these public risk assessments. This attack vector would be either a malevolent app, or a malevolent business partner of the app developer. Finally, we briefly discuss some quantitative aspects, tied to the attack surface and harms deriving from our findings.
\section{Attacks}
The attacks of concern are of two types: re-identification attacks and false positive attacks.
\subsection{Re-identification attack}
In this scenario, the attacker tries to infer who has been infected. In the GAEN scheme, Temporary Exposure Keys are used as seed to deterministically generate, together with metadata on emission power, the so-called Rolling Proximity Identifiers (RPIs) that are broadcast within each time interval. If someone is eventually found to be infected, the protocol requires to upload the TEKs for the contagious period to a central server, so they can be shared freely with all parties. From those TEKs, a long sequence of the RPIs can be generated once again, this time by an attacker. This provides new linkage information to an attacker. An attacker can then make good use of any database of historic Bluetooth beacons (since the RPI constitutes a large part of any GAEN beacon). The attacker could seek to filter out the beacons emitted by the unfortunate individual and deduce the location of the individual at any given time, or attach additional information such as a Bluetooth MAC address which is emitted simultaneously with the Bluetooth beacon (and tied to a vast ecosystem of online surveillance). If this is done sufficiently many times, it can lead to re-identification of the individual through correlation attacks with side databases (camera recording in the street, payment information, etc). One proof of concept of such an attack on DP-3T, the precursor system to SwissCovid, has been published before \cite{Otto}, and SwissCovid remains vulnerable to such an attack (if not even more so). 

\subsection{False positive attacks} 
In this scenario, the attacker can compromises the integrity of the system by
creating false contact events among users. The presumed goal is to create false
positives in risk notifications and so the attack is accomplished if the fake
contact events later correspond to positive diagnoses.

False positive attacks are achieved by either relaying (so-called \emph{Lazy Student} attack, who tries to get his class cancelled by falsely notifying the whole group sitting in the same room for a while \cite{DP3T}) or
replaying beacons previously heard. The design of the contract tracing system
means that there is a limited time when beacons can be replayed before they
become invalid.

The attacker can
increase their chance of succeeding in the attack by harvesting beacons
near locations such as COVID-19 testing sites or treatment wards (so-called \emph{Hospital
Replay} attack).

The report by Vaudenay and Vuagnoux makes two relevant observations to this type
of attacks as it relates to SwissCovid\footnote{In fact, these weaknesses affect all GAEN-based contact tracing systems.}.

\begin{itemize}
\item While any particular RPI is only broadcast for 10 minutes by the particular smartphone that generated it, the window of time during which a particular RPI can be productively replayed is quite large: two hours\footnote{Our understanding is that there is a trade-off here between privacy and security of the system. One of the goals seems to be to avoid having to store on the phones for an extended period of time (two weeks) a precise log of each beacon heard alongside a precise timestamp. This might possibly be for security reasons when confronted to an adversary with physical access to the phone. We observe that this is the result of a joint decision of Apple and Google, performing world-scale arbitrage on the tension between competing goals.}. This is concerning because it corresponds to the window of time that an attacker can mount a replay attack of RPIs that are observed.
\item The so-called \emph{Associated Encrypted Metadata} (AEM) can be tampered with. This metadata contains the Bluetooth emission power, which is deemed necessary for the distance calculation as it affects the Bluetooth attenuation calculation. This data is merely encrypted with AES-CTR, not authenticated. As a result one can tamper blindly with those values by flipping bits. This would result in false inference by the receiving device of the beacon attenuation and thus the distance traveled by each individual Bluetooth beacon\footnote{To be very specific, this inference only occurs much later, only if the (original) emitter of the beacon is declared as infected. Indeed, the encrypted metadata cannot be decrypted without computing the \emph{Associated Encrypted Metadata Key}, itself derived from the \emph{Temporary Exposure Key}, whose reveal is what represents a COVID-19 positive diagnostic.}. Since there are many obscure elements in the GAEN framework, especially at the Bluetooth level, we cannot exclude some strategy by the attacker to help them achieve their objectives more deliberately. In fact, the first author suspects some strategies do exist.
\end{itemize}

\subsection{Attack vector}
The two attacks presented above are described by the time the
attacker is listening in or replaying specific beacon payloads. It is worth
considering how the attacker would get themselves in a position to execute
these last attack steps.

One scenario would be that they are themselves deploying
hardware as a vector to the attack, or that they compromise the security of devices
to infect them with malware to mount the attack. Given our knowledge over the advertising
ecosystem, however, we find it much more credible that an attacker would implement this
attack by embedding the necessary code in an advertising or monetization
library that is sold or made accessible to app developers. In this way, the
attacker recruits oblivious end-users as vehicles for the attack.

Attackers can target their monetization SDK to apps that already hold the
relevant permission and are widely used in a target area in order to reduce the
cost of mounting the attack. Attackers could also
pay or coerce library developers to insert such vulnerabilities in the
resources they make available. This is well in reach for a state-level
adversary who wishes to inflict economic pain or confusion on a foreign country, and matches
techniques previously deployed by such actors. 

We illustrate the feasibility of
this attack with an example of a mobile app used by millions that already
harvests data from Bluetooth beacons. It would need only a small change to
the existing code to implement the attack we describe, and it can be done
entirely extrajurisdictionally.

\section{Perspective of Swiss authorities}
\subsection{National Cyber Security Centre}
As explained earlier, the Swiss National Cyber Security Centre issued a report \cite{NCSC}. We now quote a few extracts from this report, specifically on the re-identification attack through eavesdropping. 
\begin{framed}
\begin{quoting}
As one researcher pointed out, there exists an additional attack vector by using existing collection of BLE beacons where an attacker might combine these with information gained elsewhere. There are several organizations that collect Bluetooth beacons in order to geolocate users. One good example are Facebook's location tracking or several SDKs that are being used for advertisements.

[..]

Another and maybe even more serious threat could originate from organizations that operate stationary tracking systems such as payment providers relying on Bluetooth or WIFI operators. There is no real safeguard with the current design against this attack vector, \textbf{however there are only few operators of such systems and they are under the Swiss jurisdiction which gives at least some protection on the legal level}.

[..]

The public should be informed that people can turn on and off the app at any time and so stop broadcasting RPIs for defined periods of time. It is important to keep the app running whenever infection situations with unknown people can occur, but it is better to turn it off at home, which reduces the replay attack risk on the receiving side, when in places that should later not be exposed, or when at work if a risk of BLE collectors operated by the employer exists. Using the app is not a binary decision, but can be adapted by users depending of their current environment.
\end{quoting}
\end{framed}
We do not fundamentally oppose the quotes above, except for the relative ranking of the two threats (with weak commitment from NCSC to that ranking). We observe that there is only a reference to Swiss law in relation to \emph{stationary} tracking systems.

Additionally the report mentions:
\begin{framed}
\begin{quoting}
It is difficult to estimate the likelihood that someone really tries to poison the whole system by large-scale replay attacks or to try to put individuals wrongly into quarantine by a targeted re-play attack against some persons. As notified individuals are now eligible by law to free testing, targeted attacks would only be effective for a short time (until the results of the tests are known). Large-scale replay attacks would generate a high number of tests and thus be easy to detect. In case the worst case, the app could then be disabled. The damage would be limited to development costs of the app, and a lost opportunity. The alternative would be not do offer a proximity tracing app at all – even an app using a centralized approach would be vulnerable to replay attacks to a certain, though lower, degree.
In the end, this is a question that cannot be answered on a technical level alone. \textbf{While such an attack is possible, it is associated with high costs for the attackers}. As such, we believe that the benefit of the app is higher than the potential risk of such an attack.
\end{quoting}
\end{framed}

We dispute several assertions in this quote. 
\begin{itemize}
\item We agree that it is difficult to estimate the likelihood that someone really tries to poison the system. However a starting point would be to assess the technical difficulty of doing so. By showing an example here (under specific constraints), we hope to move this discussion forward a bit. While the example shown here might appear very random and small scale, our quantitative discussion at the end will show that relatively small actors could actually be impactful. State-level adversaries would definitely have the possibility to acquire the technical capacity to create havoc. 
\item We do not think the worst case would necessarily be so clear. An attacker with this type of leverage over the system will not necessarily use it right away and overtly. Instead, in certain disinformation attacks, the attacker would work at changing the perception of the system by different actors, thereby polarizing people around it. We have specific scenarios in mind of how to achieve this, but this is not the focus of the present report.
\item We contend that the cost of technically performing the attack could be
negligible to some actors, and we show this with an example of a mobile app SDK
that already harvests Bluetooth beacon data from millions of users and sends it to central servers.
\end{itemize}

\subsection{Federal Office of Public Health}
The Swiss Federal Office of Public Health (OFSP in French, BAG in German) also issued their own report \cite{FOPH}. 

\begin{framed}
\begin{quoting}
Both the original  \guillemotleft Replay attack\guillemotright ~and the variant with AEM-tampering \textbf{require that the attackers have devices that are physically in Switzerland and in proximity of to-be-tested-positive persons as well as victims for at least 15 minutes continuously. As such, these attacks cannot be conducted as remote Cyberattacks from different neighbourhoods, cities or from a foreign country}. Both attacks may be punishable under the Swiss Penal code. For example, AEM tampering may be punishable as damage to data under article 144bis paragraph 1 of the Swiss Criminal Code.
\end{quoting}
\end{framed}

We disagree with a few points presented in the quote above from the FOPH, which
motivated preparing this report:
\begin{itemize}
\item The beacons can be harvested within milliseconds because they only need to
be heard once. Legitimate devices will also log identifiers heard briefly, but only act further on those identifiers they were exposed to for a long time. An adversary would obviously not have this limitation, and can exhibit arbitrary behaviour such as relaying identifiers after first hearing their transmission over a public Bluetooth broadcast.
\item \textbf{The attacker (singular) need not be present in Switzerland, but
	merely needs to have control over devices that are physically in
	    Switzerland. This is in direct contradiction to the explicit and crucial assertion in the quote above. }
\item The AEM tampering variation helps as it expands the range of hardware that
could be leveraged, to include commodity smartphones. More precisely, it makes
smartphone-based attacks more cost effective than attacks leveraging dedicated
hardware, as the beacons can be emitted at fairly normal power but simply lie
about that emission power. The reach will not be as effective as strong
signals sent with dedicated hardware, but the trade-off would be worth it to
some types of attackers who can control a large number of smartphones. If the hardware offers the possibility of broadcasting signals at higher power, this feature could definitely be used in conjunction with the AEM tampering.
\end{itemize} 

The first and third points need not be discussed more in this work but rather we
focus at length on the second point (in bold). Before that we include for
reference the relevant article from the Swiss Criminal Code below. 

\begin{framed}
\begin{quoting}
 Art. 144bis
 
 1.  Any person who without authority alters, deletes or renders unusable  data  that  is  stored  or  transmitted  electronically  or  in  some  other  similar way is liable on complaint to a custodial sentence not exceeding three years or to a monetary penalty. If the offender has caused major damage, a custodial sentence of from one  to  five  years  may  be  imposed.  The  offence  is  prosecuted  ex  officio. 

2. [..]
\end{quoting}
\end{framed}

While we are not lawyers, we additionally observe:
\begin{itemize}
\item The reference to 144bis for the AEM-tampering attack is fairly speculative, as this attack requires no interference with someone else's transmission but instead just passive listening and then separately replaying. 
\item No article of law is referenced for the mere replay attack, which does not actually require anything but re-emitting packets on the Bluetooth spectrum, within the ISM radio bands\footnote{\url{https://en.wikipedia.org/wiki/ISM_band}}. Note however that it is possible that members of the Bluetooth Special Interest Group have mutual contractual obligations to protect the usage of specific service UUIDs over the Bluetooth spectrum. We observe that Apple bought or leased the UUID relevant for the GAEN framework, \textsf{0xFD6F}\footnote{see the list of service UUIDs at  \url{https://www.bluetooth.com/specifications/assigned-numbers/16-bit-uuids-for-members/}}.
\end{itemize}

\section{Smartphone-based Attack}
We contend that the arguments outlined by Swiss authorities that this attack is
infeasible or costly are based on the incorrect mental model of an attacker
listening for Bluetooth signals in one location, traveling to another, and
rebroadcasting them.  Instead, we propose a simpler attack in which a
third-party library, which we call the \emph{attack SDK}, is included
in a popular benign-looking app, and this library implements the relay attack.
As such, end-users act as \emph{confused deputies} unknowingly undermining the
integrity of the contract tracing system.

\subsection{Attack Description}
The attack works by having an attack SDK that uses the Bluetooth scanning feature to listen for all
observed RPIs. As soon as one is observed, the attack SDK uploads the RPI along
with geolocation to the attacker. The attacker may then propagate the RPI to
other mobile devices running the same attack SDK, and these other devices then
start to broadcast the RPI. Devices operating normally nearby will observe these RPIs
and incorrectly conclude that they are in close proximity to the original
broadcaster of the RPI. This attack could be amplified by tampering with the AEM.

The implementation of a two-hour window for RPI reuse makes this attack easier
to mount as there are fewer RPIs to collect and transmit. Despite that, however, a much
shorter window does nothing to prevent the feasibility of this attack. As soon
as an RPI is observed, it can be sent over the Internet to the attacker and
operationalized on other devices. This is an unavoidable consequence as long as
the underlying distance-bounding scheme is vulnerable to a relay attack. Such an attack has actually been performed by security researchers, on the pre-GAEN version of SwissCovid \cite{GAP}.

The attacker may use the geolocation information associated with the RPI to
select the particular RPIs to rebroadcast. For example, a state-level adversary
wishing to inflict economic hardship on a foreign country may want to shut down
particular industries by making it appear as though there is an outbreak of
disease at such sites. They may use a strategy of collecting RPIs observed near
hospitals or testing facilities, and rebroadcast the RPIs at work sites
relevant for the industry.

Observe that the attacker described need have no relation, legal or otherwise,
with the country in which the attack is mounted. Neither the attack SDK nor the
benign-looking app need to have any connection with the affected country. It is
only the end users unknowingly undermining the system that have a connection to
the affected country.

This is not an exotic attack that requires a high-level of sophistication or
domain-specific knowledge. The implementation effort of scanning and
broadcasting Bluetooth signals is greatly reduced with APIs designed exactly to
facilitate this kind of development. Furthermore, there is no need for user's
phone to be infected with malware for this attack to succed. Instead users
willingly download an app that contains the attack SDK and are unaware that they
are participating in this attack.

The app maker can be incentivized to include this SDK by paying them based on
their install base. Indeed, this is the primary means of app monetization.
There is already ample evidence of a rich ecosystem of surveillance on mobile
devices in the forms of ads and analytics platforms that allow app developers to
be monetized based the usage of their apps~\cite{coppa}. Later in this section we show
the existence of a SDK with millions of installations that already harvests MAC addresses and sometimes
advertising data from Bluetooth beacons.

\subsection{Smartphone Permissions}
The Android platform\footnote{We focus on Android smartphones as there will be a long tail of Android phones able to carry such attacks, due to longer and more complex cycles of updates as well as a larger variety of operating system providers. On iOS, as part of recent upgrades, the OS introduces a new filter for Bluetooth packets with service UUID \textsf{0xFD6F}, making apps oblivious to SwissCovid beacons} uses a permission system to protect access to sensitive
resources based on the security principle of \emph{least privilege}.
Apps must request relevant permissions in order to be able
to perform specific functionality. For Bluetooth, there are two relevant
permissions: \textsc{bluetooth} and \textsc{bluetooth\_admin}. The former allows
the use of Bluetooth devices, while the latter permits scanning for and connecting
to new Bluetooth devices.

Since Android 6.0, apps are required to hold a location permission
in order to scan for nearby Bluetooth devices, as well as scan for nearby WiFi
routers. This is because the serial numbers for these devices can act as a surrogate for
location~\cite{wifileaks}. Consequently, the Android
permission system now considers the collection of nearby device's MAC addresses to require a location permission as well.

Crucially, the permissions that an app has access to are also granted to any
other code that runs as part of that app. This includes the prolific ads and
analytics networks that are often included for monetization of apps. These can
report back user data to servers and offer other features to developers. The use
of such third-party code is common---it is easier and generally better to make
use of existing reusable code than to rewrite everything \emph{ad hoc}.
Nevertheless, it means that some third-party code can find itself embedded in an
app that has requested the correct permissions and is  installed on millions of
devices. This is the mechanism through which we see the hospital relay attack being most
easily mounted. Note that this can be done entirely extrajurisdictionally.

Note that despite the significance of having administrative power over Bluetooth on a
device, it is by no means a rare permission to hold. Android does not even
consider it a ``dangerous'' permission, that is, one for which users must be
further asked when it is first used.  From a random selection of
1550 apps\footnote{We built a list of more than 50000 apps by searching Google
	Play Store for
	random adjectives, shuffling the results, and downloading the first
	    1550.},  129 hold the \textsc{bluetooth} permission, 81 hold the
	    \textsc{bluetooth\_admin} permission, and 63 (4\%, or 1 in 25) hold both Bluetooth
	    permissions as well as \textsc{access\_fine\_location}.
We next illustrate the results of an analysis of one of these 63 apps.

\subsection{PixelProse SARL}
PixelProse SARL\footnote{SARL stands for \emph{Soci\'et\'e \`a responsabilit\'e
	limit\'ee}, i.e., Limited Liability Company} is listed as a developer on
	the Google Play
	Store\footnote{\url{https://play.google.com/store/apps/developer?id=PixelProse+SARL}},
	with id \textsf{PixelProse+SARL}. Judging from the additional
	information entered in the Play Store, the business register and the
	website hosting the privacy policy of their apps, it looks to actually
	be a one-man operation located in a village located a mere 25\,km outside of Switzerland, near Annecy, France.

\begin{figure}[h!]
\centering
\includegraphics[scale=0.25]{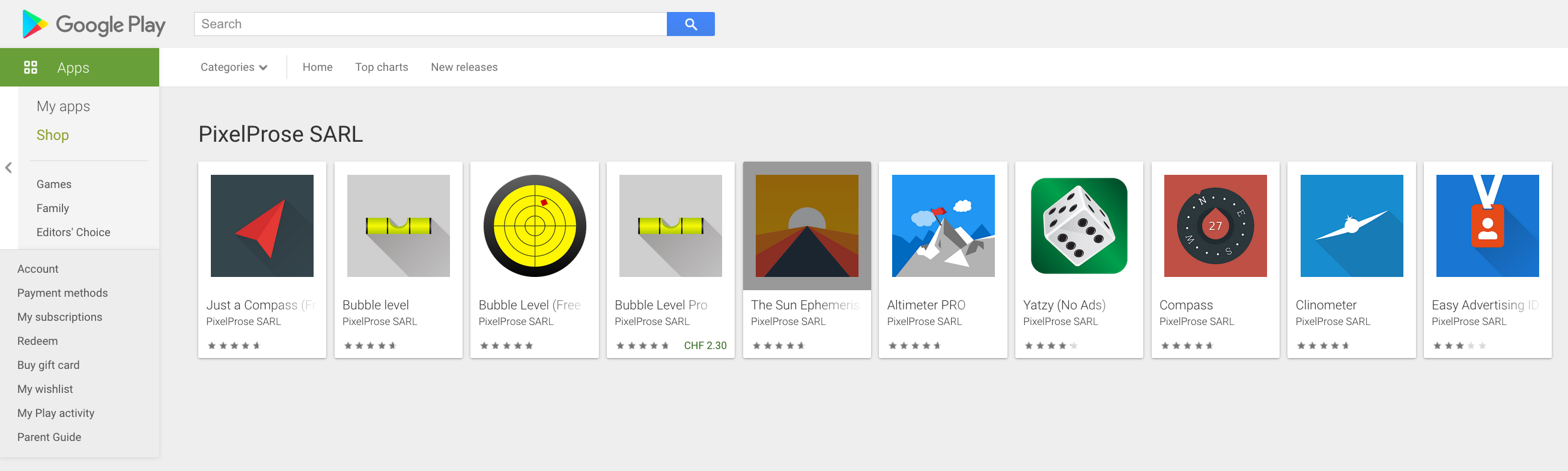}
\caption{The apps listed on the Google Play Store by PixelProse SARL.}
\end{figure}

\begin{center}
\begin{table}
\begin{tabular}{p{30mm}ccccc}
\toprule
\textbf{app} & \textbf{id} & \textbf{installs} & \textbf{ reviews} &
\textbf{rating} \\
\midrule
Bubble level &
\textsf{net.androgames.level}&
10M+&
218 401 &
4.7\\
\midrule
Compass&
\textsf{fr.avianey.compass}&
1M+&
36 801&
4.6\\
\midrule
Just a Compass (Free \& No Ads) & 
\textsf{net.androgames.compass} &
1M+ &
21 711 &
4.7\\
\midrule
The Sun Ephemeris (Sunset, Sunrise, Moon position)&
\textsf{fr.avianey.ephemeris}&
50 000+ &
1 013&
4.6\\
\midrule
Altimeter PRO&
\textsf{fr.avianey.altimeter}&
50 000+&
1 008&
4.6\\
\bottomrule
\end{tabular}

\caption{Google Play Store statistics on the PixelProse SARL apps requesting \textsc{bluetooth\_admin} and location permissions.}
\end{table}
\end{center}

On the Google Play Store, PixelProse SARL is listed as developer for ten apps. Most have basic overt functionality, leveraging one of the phone sensors (level, altimeter, compass, etc.). One app simply exposes the advertising identifiers associated to the phone (and requires no permission). Such apps are often used and offered by players in the adtech industry as a debug tool for their systems. 
Five of the apps are notable because they request \textsc{bluetooth\_admin} and location permissions.

\subsection{Experimental Testbed} 
Our dynamic analysis testbed consists of a Pixel 3 mobile phone. It is running an instrumented version
of Android Pie that collects all network traffic, including
traffic secured by TLS. Our instrumentation attributes all network traffic to the
specific app that is responsible for its transmission. As such, we are able to
observe the actual real-world behaviour of an app as it executes.

Our instrumented operating system further injects spurious Bluetooth scan
results to the app in question. We use conspicuous palindromic MAC addresses in our injected
results and search for them being transmitted. We format the Bluetooth
advertising data to match standard beacon formats. For example, we have an
iBeacon (MAC address \texttt{AB:B1:E6:6E:1B:BA}), an AltBeacon
(MAC address \texttt{AB:B1:E7:7E:1B:BA}), an Eddystone URL (MAC address
\texttt{AB:B1:E8:8E:1B:BA}), and a GAEN beacon (MAC address
\texttt{AB:B1:E9:9E:1B:BA}).

We start the app and accept its request for permissions, and then leave
the app running without disturbing it.
We then obtain the network traffic and process it with a suite of decoders to
remove standard encodings such as gzip and base64. We remove network traffic
from other apps, such as system ones, and consider only that traffic being sent
by the app under investigation.

\subsection{Case Study: Bubble Level}
In this section we discuss the findings for one app, PixelProse SARL's Bubble
level (\textsf{net.androgames.level}). This app is notable as it has more than ten million installations (but less than fifty million) and more than 200,000 reviews. While other apps from the same company exhibited
the same or similar behaviours, we focus on this one for our case study.

On June 16th, 2020, we downloaded and installed the app
\textsf{net.androgames.level} from the Google Play Store to our instrumented Pixel 3 mobile
phone and ran it with our dynamic analysis testbed.

It first contacts
\texttt{bin5y4muil.execute-api.us-east-1.amazonaws.com}
(port 443) where it performs a GET request for \texttt{/prod/sdk-settings}. It
returns a JSON object storing a configuration. This includes a number of
parameters for Bluetooth scanning:
\begin{packed_item}
    \item \texttt{"baseUrlDomain":"api.myendpoint.io"}
    \item \texttt{"beaconsEnabled":true}
    \item \texttt{"bleScanMaxPerHour":2}
    \item \texttt{"btScanMaxPerHour":2}
\end{packed_item}

We also found the app sending the unique advertising identifier to the following domains:
\begin{packed_item}
\item  \texttt{adjust.com}
\item  \texttt{adsmoloco.com}
\item   \texttt{cuebiq.com}
\item    \texttt{doubleclick.net}
\item    \texttt{kochava.com}
\item    \texttt{mobvista.com}
\item    \texttt{mopub.com}
\item    \texttt{myendpoint.io}
\item    \texttt{smartechmetrics.com}
\end{packed_item}

The most interesting results are the transmissions to 
\texttt{smartechmetrics.com} and
\texttt{myendpoint.io}, which include full scans of nearby wireless
devices. Tables~\ref{smartechmetrics}--\ref{myend-last} show example
transmission. Both domains receive the results of a
scan of nearby WiFi routers and Bluetooth devices as well as precise
geolocation. 

Table~\ref{smartechmetrics} presents an observed transmission to
\texttt{smartechmetrics.com}.
We have redacted identifying information, and observe that some of
the transmissions of WiFi routers are devices that are
nearby to the testbed but not actually devices owned by the authors. We see that
the MAC addresses for all injected Bluetooth traffic are collected and
transmitted, along with two pieces of consumer Bluetooth electronics in the same
room as the testbed.

\begin{table}
\footnotesize
\begin{verbatim}
{
  "obs": [
    {
      "gaid": "XXXXXXXX-XXXX-XXXX-XXXX-XXXXXXXXXXXX",
      "ids": [  "google_aid^XXXXXXXX-XXXX-XXXX-XXXX-XXXXXXXXXXXX" ],
      "lat": 51.XXXXXXX,
      "lon": -114.XXXXXXX,
      "metadata": [
        "device:AOSP on sargo",
        "os_version_int:28",
        "sdk_version:1.9.2-bcn",
        "app:Bubble"
      ],
      "observed": [
        {
          "mac": "AB:B1:E9:9E:1B:BA",
          "name": "null",
          "rssi": -12,
          "tech": "ble"
        },
        {
          "mac": "AB:B1:E7:7E:1B:BA",
          "name": "null",
          "rssi": -12,
          "tech": "ble"
        },
        {
          "mac": "AB:B1:E6:6E:1B:BA",
          "name": "null",
          "rssi": -12,
          "tech": "ble"
        },
        {
          "mac": "AB:B1:E8:8E:1B:BA",
          "name": "null",
          "rssi": -12,
          "tech": "ble"
        },
        {
          "mac": "XX:XX:XX:XX:XX:XX",
          "name": "XXXXXXXX",
          "rssi": -40,
          "tech": "wifi"
        },
        {
          "mac": "XX:XX:XX:XX:XX:XX",
          "name": "XXXXXXXX",
          "rssi": -60,
          "tech": "bluetooth"
        },
        < truncated >
      ],
      < truncated >
    }
  ]
}
\end{verbatim}

\caption{Data sent by \textsf{net.androgames.level} to
	\textsf{api.smartechmetrics.com}. Note that \texttt{truncated} denotes truncated data,  \texttt{XXXX} denotes
redacted values, and that we added whitespace to improve clarity.}
\label{smartechmetrics}
\end{table}

\begin{table}
\footnotesize
\begin{verbatim}
[
  {
    "beaconType": "EDDYSTONE_URL",
    "isCharging": false,
    "loc_at": 1592412157818,
    "mac": "AB:B1:E8:8E:1B:BA",
    "name": "",
    "rssi": -12,
    "scan_record": {},
    "tech": "ble",
    "time": 1594004410113
  },
  {
    "isCharging": false,
    "loc_at": 1592412157818,
    "mac": "AB:B1:E7:7E:1B:BA",
    "name": "",
    "rssi": -12,
    "scan_record": {},
    "tech": "ble",
    "time": 1594004410112
  },
  {
    "beaconType": "IBEACON",
    "isCharging": false,
    "loc_at": 1592412157818,
    "mac": "AB:B1:E6:6E:1B:BA",
    "name": "",
    "rssi": -12,
    "scan_record": {},
    "tech": "ble",
    "time": 1594004410112
  },
  {
    "isCharging": false,
    "loc_at": 1592412157818,
    "mac": "XX:XX:XX:XX:XX:XX",
    "name": "XXXXXXXXXXXXXXXX",
    "rssi": -53,
    "scan_record": {},
    "tech": "bluetooth",
    "time": 1592412158800
  },
  {
    "isCharging": false,
    "loc_at": 1592412157818,
    "mac": "XX:XX:XX:XX:XX:XX",
    "name": "XXXXXXXXXXXXXXXX",
    "rssi": -54,
    "tech": "wifi",
    "time": 1592412158401
  },
      < truncated >
]
\end{verbatim}
\caption{Data sent by \textsf{net.androgames.level} to
	\textsf{api.myendpoint.io}.}
\label{myend-scan}
\end{table}

\begin{table}
\footnotesize
\begin{verbatim}
{
  "all_beacon_data": [
    {
      "accuracy": 20.934XXXXXXXXXXX,
      "ad_id": "XXXXXXXX-XXXX-XXXX-XXXX-XXXXXXXXXXXX",
      "altitude": 1052.XXXXXXXXXX,
      "beacons": [
        {
          "distance": 0.97085,
          "layout_name": "",
          "mac_address": "AB:B1:E6:6E:1B:BA",
          "major": "53479",
          "minor": "42571",
          "mumm": "AB:B1:E6:6E:1B:BA_01022022-fa0f-0100-00ac-dd1c6502da1c_53479_42571",
          "rssi": -12,
          "uuid": "01022022-fa0f-0100-00ac-dd1c6502da1c"
        }
      ],
      "bearing": 0,
      "latitude": 51.XXXXXX,
      "longitude": -114.XXXXXXX,
      "model": "AOSP on sargo",
      "os_version": "9",
      "platform": "android",
      "sdk_version": "1.9.2-bcn",
      "speed": 0,
      "time": 1592412157818,
      "vert_acc": 2
    }
  ]
}
\end{verbatim}
\caption{Data sent by \textsf{net.androgames.level} to
	\textsf{api.myendpoint.io}. Observe that this includes the
	    transmission of the advertising data from a Bluetooth
	    beacon.}
\label{myend-last}
\end{table}

Table~\ref{myend-scan} shows an example WiFi and Bluetooth scan being
send. We see that in addition to actual wireless devices, all of the spurious
Bluetooth devices are included in the transmission. Note also that there is a
\texttt{scan\_record} field, which is set to an empty JSON object.  We looked at
the initial configuration that the app receives and found a
\texttt{"shouldAddScanRecord":true} as a parameter; we have reverse engineered
the app but did not determine the conditions that may make this field store more
data.

Table~\ref{myend-last} shows another interesting transmission sent to \texttt{api.myendpoint.io}. This
includes not only precise geolocation and a result from a Bluetooth scan but it actually
includes the \emph{advertising data} of the Bluetooth scan itself.  The list of beacons in the
JSON transmission has one entry corresponding to the injected iBeacon. The key
\texttt{mumm} we believe stands for \emph{mac uuid major minor}, because it is
in fact an underscore-separated string of those four fields. We observe that the
uuid, major, and minor values are exactly those that we configured to be sent as the
advertised data.

From this case study of \emph{a single app}, we understand that it is
\emph{already the case today} that Bluetooth advertising data is being read,
processed, and sent to servers on the Internet by the \emph{millions of
users} of this app while they go about their day.
There is no technical limitation that prevents the
full collection of the advertised data: a few simple lines of code could make
them also upload the GAEN RPIs. These can be then sent out to other devices
and rebroadcasted.

\subsection{Privacy policy}
To their credit, PixelProse's privacy policy\cite{policy} is fairly transparent with
its practices. We reproduce a large part of it here.

\begin{framed}
\begin{quoting}
The information gathered when you interact with the application falls into Non-personal information only collected through technology, which includes tracking information collected by us as well as third parties. Applications that uses Facebook SDK features such as “login with facebook”, “share”, “like” might send additional information to facebook (see learn more section for facebook privacy policy and platform compliance). Information retrieved from Facebook are used for display only within the app (in any case, those data are sent to third party services).

WE DO NOT READ and/or BROWSE your personal data on your phone : this include PHOTOS, CONTACTS, CALL LOGS, DOCUMENTS, etc

You acknowledge that we may collects and passes to our trusted partners a variety of information for use as described below:

\textbf{What's collected}

Mobile advertising IDs (e.g., iOS IDFAs and Android Advertising IDs), precise location information (such as your device GPS coordinates), relative location information (from WiFi signals or Bluetooth Low Energy devices in your proximity, for example), device-based advertising identifiers, information about your mobile device such as type of device, operating system version and type, device settings, time zone, carrier and IP address.

Our software may use anonymous, statistical or aggregated information collected and accessed through wireless and more specifically Bluetooth radios, by transmitting or otherwise communicating or making available such information to users of its Services, to the Services' providers, partners and any other third party.

\textbf{When we collect it}

When the service is in use or running in the background.

\textbf{How we use and share it}

We and our trusted partners may use the information (a) to customize ads in this service or other services (for instance, if your device is often located at or near music venues, you might receive offers for music tickets); (b) to measure effectiveness of those ads, or (c) for market, civic or other research regarding aggregated traffic patterns (for instance, a company that analyzes shopping trends might want to learn whether more or fewer devices are seen near malls or in other shopping districts). 
\end{quoting}
\end{framed}

The terse statement about when they collect the data fails to emphasize an
important point: the app will automatically start when the device is started.
This means that it is \emph{always} running in the background unless the user
takes an unusual and deliberate action to \emph{force stop} the app through a sequence of
system dialogs. Therefore, for most end users who have installed this app, the
data collection is happening continually as long as it remains installed.

There is an explicit acknowledgement in this privacy policy that Bluetooth Low
Energy devices in proximity, geolocation, and direct advertising identifiers are being collected and shared with ``trusted partners''. In this privacy policy, there is a link leading to a new page, itself containing hyperlinks to all of PixelProse's trusted partners\footnote{\url{https://pixelprose.fr/trusted-partners/}}, see Table~\ref{partners}. A few entries in that table themselves refer to additional partners ``one hop down''. This includes for instance the entry ``Cuebiq and its Partners''. Since those partners are themselves listed on Cuebiq's website, we include them in Table~\ref{cuebiq} as well.

\begin{table}
\begin{multicols}{2}
\footnotesize
\noindent Google\\
Facebook\\
Cuebiq and its Partners\\
Sense360\\
X-Mode Social, Inc.\\
Nodle International\\
Advan Research Company\\
AdSquare\\
AirSage\\
Adobe\\
Adform\\
Amazon\\
Amobee\\
AppNexus Inc.\\
AreaMetrics\\
Arrivalist\\
AT\&T\\
AWS\\
Beaconinside\\
BDEX LLC\\
Bloomberg L.P. \\
Carbon Reach\\
Cisco Systems, Inc. (Meraki LLC)\\
Complementics\\
Conversant Europe Ltd.\\
Descartes Labs, Inc.\\
Drawbridge, Inc.\\
DataStream Group\\
Exterion\\
Equifax\\
Facebook\\
Factual\\
FourSquare\\
Freckle IoT Ltd.\\
Generali\\
GEOBLINK SL\\
GeoUniq\\
Google \\
Gravy Analytics\\
GroundTruth\\
Gyana\\
HERE Technologies\\
Hyas\\
Hyp3r\\
Infinia Mobile\\
Intersection\\
InMobi\\
JLL\\
Jorte\\
Kantar Media\\
Location Sciences\\
Locarta GmbH\\
Locomizer\\
Liveramp\\
LoopMe\\
MediaMath\\
MiQ\\
MyTraffic\\
Native Touch\\
Nodle International\\
On Device Research\\
OnSpot Data\\
OpenX Software Ltd. and its affiliates\\
O'Reilly Automotive Stores\\
Oracle\\
Pelmorex\\
Peroni\\
Pitney Bowes\\
Placed\\
Place Dashboard\\
PlaceIQ\\
Placense\\
Popertee\\
PubMatic, Inc.\\
PushSpring\\
Radiant Solutions\\
S4M\\
Sierra Nevada Corporation\\
Sito Mobile\\
Skyhook Wireless\\
Statiq\\
Systems and Technology Research\\
Talon Outdoor\\
Tamoco\\
Teemo\\
Telefonica\\
Thasos\\
Tiendo\\
The Singlespot\\
The Trade Desk, Inc and affiliated companies\\
The Rubicon Project, Limited\\
UberMedia\\
Upsie\\
Vectaury\\
Vertical Scope\\
Vistar Media\\
Wireless Registry dba SignalFrame\\
Xandr\\
xAd, Inc. dba GroundTruth\\
Zeotap
\end{multicols}
\caption{Trusted partners of PixelProse}
\label{partners}
\end{table}

\begin{table}[]
\begin{multicols}{2}
Adobe\\
Adsquare\\
Advan\\
Amobee\\
Arrivalist\\
Beintoo\\
Bridge\\
Centro\\
Clear Channel Outdoor\\
Crosswise\\
Dstillery\\
Epsilon\\
Experian\\
Fysical\\
Google\\
Groundtruth\\
Hivestack\\
Horizon Media\\
Infinia Mobile\\
Inscape\\
Intersection\\
Kantar\\
LiveRamp\\
Lotame\\
MaxMind\\
MobileFuse\\
Mogean\\
MyTraffic\\
Native Touch\\
Near.co\\
Neustar\\
Oracle\\
Pelmorex\\
PlaceCast\\
Placer.ai\\
Point Inside\\
PushSpring\\
Qualia\\
Salesforce\\
Samba TV\\
Semcasting\\
Simpli.fi\\
The Trade Desk\\
UberMedia\\
Ubimo\\
Valassis\\
Verizon Media\\
Viant\\
Wieden+Kennedy
\end{multicols}
\caption{Cuebiq's partners}
\label{cuebiq}
\end{table}

\normalsize
\subsection{X-Mode Social}
We now discuss the COVID-related activities of one PixelProse partner in particular, X-Mode Social. In an April 4th 2020 interview\footnote{\url{https://edition.cnn.com/2020/04/04/tech/location-tracking-florida-coronavirus/index.html}}, CNN journalist Donie O'Sullivan and X-Mode Social\footnote{\url{https://www.xmode.io/}} CEO Josh Anton had the following exchange\footnote{It is hard to convey through a transcript the changing tone of the journalist, sometimes talking to the CEO and sometimes to the audience. The reader is encouraged to watch the video as it is illustrated with informative graphics.}:

\begin{quoting}
    \textbf{CNN journalist:} Josh Anton runs X-Mode, a company that tracks the movements of devices like cell phones. His team says it has used location data to track where [Spring Breakers] in Fort Lauderdale[, Florida] in March went after they left. 
    
    \textbf{X-Mode Social CEO:} From New York, to the Midwest, even Canada. You know, the power of this location data - it can be used to understand not only how people and where people are traveling post-gathering, but also potentially to prepare and be proactive if something happens, to be able to identify future hotspots of where the coronavirus could happen before it happens.
    
    \textbf{CNN journalist:} But the applications of X-Mode's technology go way beyond [Spring Breakers]. Anton says the company tracks 25 million devices every month in the United States, and millions more around the world\footnote{On screen, at the same time, the text "40 millions in EU, APAC, LATAM" appears.}. X-Mode says it would be willing to work with governments and other groups to help stop the spread of the coronavirus. 
    
    \textbf{X-Mode Social CEO:} We work with apps that have a real use case for running location, whether it's transit apps, whether it's weather apps, or apps that alert you about the earthquakes happening near you. Right? We then integrate our location technology to allow data sharing, where a user can opt in to sharing their location data. We comply with GDPR, we comply with C[alifornia] C[onsumer] P[rivacy] A[ct]. [..]
    
    \textbf{CNN journalist:} The company claims it licenses that data to third parties, including advertisers, without any personal identifying information. Can your technology be used to track individuals? Can it track me? 
    
    \textbf{X-Mode Social CEO:} It could, right? But we don't allow that. And we don't allow any of our partners to do that, because we just don't think that's the right thing to do. 
    
    \textbf{CNN journalist:} While Anton says his company makes every effort to keep data secure and doesn't identify the owners of the devices it is tracking, there are serious privacy concerns about this kind of technology. In 2018, a New York Times investigation showed how location data could be used to identify the specific owner of a particular device. You know, I don't think a lot of people realize this type of technology even exists, that it's out there. And I think some people are pretty creeped out. What would you say to folks who have concerns when they sort of see that you can track devices like this? 
    
    \textbf{X-Mode Social CEO:} I am going to quote Uncle Ben from "Spider-Man," which is "with great power comes great responsibility." Right? You know, I think there is a fine line. I think it is very important that users consent to this. And it's very important that you act ethically with that data.  
\end{quoting}

In addition to this interview, X-Mode Social communicates heavily on the possibilities of this data to fight COVID-19. 

It promotes several datasets on the Amazon Data Exchange, including one titled \emph{Global (excluding EU) COVID-19 Daily Geolocation Data for Research}\footnote{see \url{https://aws.amazon.com/marketplace/seller-profile?id=4e8835bd-89dc-4ae7-818d-52c22fedcbb9} and \url{https://www.xmode.io/location-data-for-the-cure-x-mode-makes-free-data-set-available-to-covid-19-researchers/}}, and provides detailed analysis of foot traffic in grocery stores, gyms, etc., in relation to lockdowns\footnote{\url{https://www.xmode.io/location-data-in-action-covid-19-impacts-foot-traffic-in-grocery-stores-gyms-and-more/}}.

\begin{figure}[]
    \centering
    \includegraphics[scale=0.33]{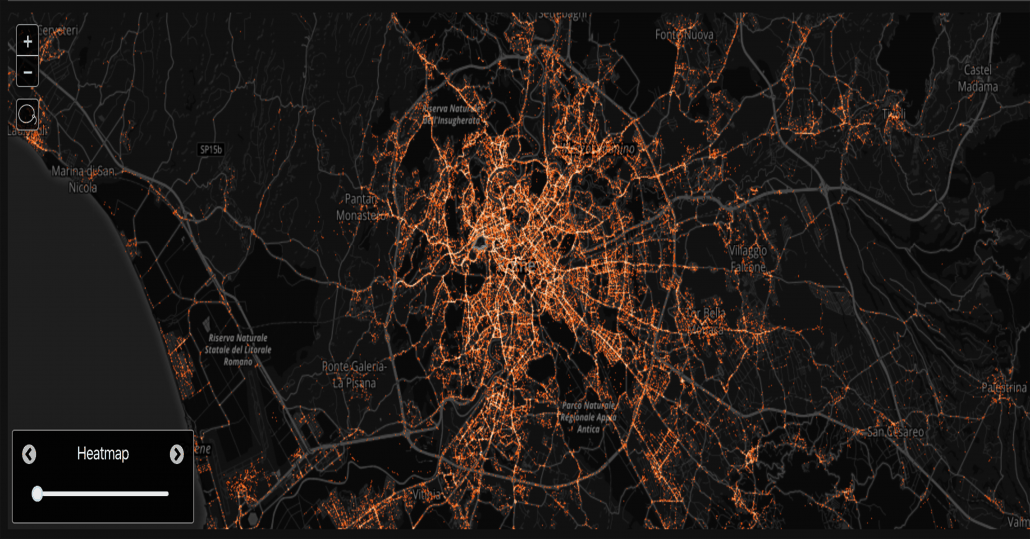}
    \caption{A heatmap of low velocity movements in Rome during lockdown (week of February 23rd 2020) \cite{rome}}.
    \label{rome}
\end{figure}

Finally, in a blog post titled \emph{Location Data in Action: Heatmaps Help Track Coronavirus Across the Globe}, X-Mode Social includes several heatmaps of Rome during lockdown (see Figure~\ref{rome}).

\section{Discussion}
\subsection{Legal considerations}
To be clear, we make no assertion here that PixelProse SARL, any of its trusted
partners, or X-Mode Social's activities violate any laws, plan to do so, or even
plan to repurpose the data they have collected anywhere for any COVID-related
purpose. We use this as an example to show \emph{an existing} network of
surveillance of Bluetooth data that uses end-users smartphones as a vehicle to
collect and transmit this data.

In particular, this example provides evidence against the assertion by Swiss authorities that the Swiss
Criminal Code would be helpful in addressing the situation of attackers
performing replay or reidentification attacks against users of SwissCovid. It is
possible to collect and rebroadcast this data using millions of smartphones
simply by providing a third-party library to app developers and incentivizing
them to include it in their apps. This is well within the abilities of a variety
of adversaries.

\subsection{Linking Identifiers}

In the GAEN protocol, the Bluetooth MAC address gets rotated to a new value along with
the RPI. This is necessary because otherwise RPIs would be trivially
linkable by looking at the corresponding MAC address. Despite that, sending both
pieces of information creates a risk of correlated pieces of data that have
entirely separate purposes.

On their own, neither the MAC address nor the RPI constitute directly identifiable data ---and hence personal data---due to their ephemeral
nature. This provides plausible deniability to any entity processing these
ephemeral values against the claim that they would be processing personal data.

Despite that, we observe that the linkage between the two types of
identifiers (Bluetooth MAC addresses and GAEN RPIs) is vulnerable to
correlation attacks, since they evolve simultaneously. This does not---on its own---constitute
a re-identification attack. It does, however bring risks: RPIs may later
be related to the \emph{public broadcast} of sensitive medical
information with the \emph{expectation} that it remains entirely anonymous.
At the same time, correlated MAC addresses may be collected by the
existing widespread ecosystem of online advertising and profiling. These two
correlated identifiers serve different purposes and it violates security design
principles to staple on a security purpose to something like a MAC address when
it was never designed with that purpose in mind.

The combination of chronology and linkage is particularly troublesome. Were a
database of MAC addresses and RPIs created, then combining it with
one of many databases storing some combination of persistent identifiers, location information,
and MAC addresses would be devastating for personal privacy. The ability to collect both
MAC address and RPIs is technically trivial, particularly were it curated
through millions of confused deputies by an attack SDK
that we described. The security argument that ensures anonymity in this
system cannot be simply ``no one would build such a database.''

Indeed, such a database could operate as an API that turns Bluetooth MAC address
to a persistent identifier like an advertising ID. Google, for example, already
offers a paid API service that turns router MAC addresses into GPS
coordinates~\cite{geolocate}. Facebook offers via its SDK a service available to apps, under the \textsf{newCurrentPlaceRequest} API, turning ambient signals (including Bluetooth beacons) into place information (and additionally gives confidence levels on multiple places, facilitating re-identification attacks confronted with noisy data) \cite{facebookapi}. This possibility could be exploited by developing an attacker-controlled app, installed on an attacker-controlled device, and broadcasting target Bluetooth MAC addresses to the attacker-controlled device.

Such an API can exist \emph{entirely outside} of the
context of Bluetooth-based proximity and contract tracing apps, though its
customer base may primarily consist of those performing de-anonymization
attacks. Google now disallows API access to the Bluetooth MAC address, but it is
a fundamental design principle in security to not staple on security purposes to
things that were not originally designed to have one.

To give further perspective on the size of the problem, PixelProse SARL is just one example plucked from an ecosystem of pervasive surveillance. For instance, the Android Beacon Library \cite{ABL}, which is a
well-liked library used by app developers, claims on its homepage to be present
in 16000 mobile applications, with 350 million installations in total (the
statistics are provided as of September 2018).  It is evident that not all apps using the Android Beacon Library are malevolent, but this should give a sense of the scale of the potential problem. 

Additionally, we would like to warn that we
have found other situations of concern, that would be relevant in a broad range
of countries, including Switzerland. Some of the apps we found had hundreds of
millions of installs, although in our experience one should probably be more worried with apps flying a bit more under the radar but still with 10M+ of installs. Indeed those tend to themselves get less investigative coverage than the bigger ones, but end up using external SDKs who do end up aggregating data from more sources and contexts, with even more opacity and are even more dependent business-wise on escalating a privacy-invasive model.

\subsection{Asymmetric Utility Growth}
Any complex system such as SwissCovid interacting with potential attackers should be the subject of extensive quantitative modelling. In particular, efforts should be made to understand sudden transitions on how this system could evolve over time, as a function of infection rates or app prevalence.

We will not go into such a detailed quantitative analysis here. We will simply mention some factors that should give pause to anyone with a quantitative background who seeks to evaluate the wisdom of deploying such a system, without engaging in more precise quantitative analysis. 

\begin{figure}[]
    \centering
    \begin{framed}
    \includegraphics[scale=0.51]{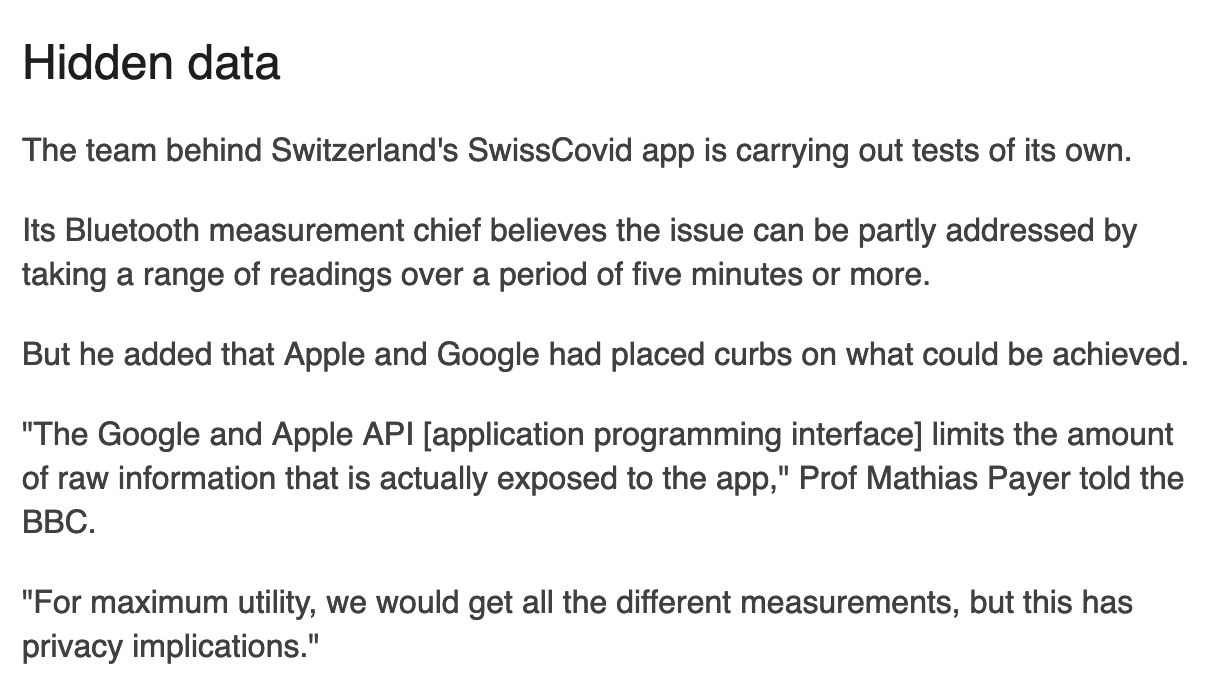}
    \end{framed}
    \caption{A discussion of the limitations introduced by GAEN \cite{BBC}.}
    \label{BBC}
\end{figure}

The utility of SwissCovid for the public health authorities is limited by the requirement that both sides install the app\footnote{Note that this is purely at the discretion of Google and Apple, as they could activate (part of) their API by default.}.  In other words, if we call the prevalence of the SwissCovid app in the population $\alpha_{SC}$, the utility of SwissCovid at detecting at-risk contact events grows (at best) like $\alpha_{SC}^2$ (\emph{i.e.,} slowly since that proportion is below 1). Additionally, as acknowledged by a DP-3T expert to the BBC, Apple and Google introduce limitations to the way Bluetooth data can be leveraged to infer distance (see Figure~\ref{BBC}). 

We now consider an attacker in the position of PixelProse SARL or one of their
trusted partners (like X-Mode Social), which would be aggregating data from
several apps. We claim this attacker could gain visibility and understanding
much faster than an app like SwissCovid of the epidemiological situation by
leveraging both the information SwissCovid broadcasts and information it
collects from the mobile devices running the attack SDK.

We divide the population into the following three categories. Note that the
three categories may intersect.
\begin{description}
\item[infected group: ] simply the set of individuals who are infected;
\item[SwissCovid group: ] the set of individuals who have SwissCovid installed;
\item[confused deputies group: ] the set of individuals whose Bluetooth and
location data is accessible to the attacker, through one or more apps or SDKs.
\end{description}
In parallel with notation previously introduced for app prevalence, we define $\alpha_{CD}$ to be the proportion of the population in the confused deputies group. 

In the suppression phase of an epidemic (and neglecting potential immunity
effects), it looks like a dynamic
understanding of the epidemiological situation is mostly derived from three
considerations: which individuals are infected, what constitutes an
epidemiologically relevant contact, and which individuals are therefore at risk.
We think that in some cases an attacker could be better placed than public
health authorities (together with individuals in the population) to infer these characteristics. In other words, the utility would grow faster for an attacker than it would for individuals and public health authorities together. 

\begin{description}
\item[close contacts:] As suggested in Figure~\ref{BBC}, having the freedom to
scan or broadcast at the Bluetooth layer gives more precision on distance
inference from Bluetooth than the GAEN framework affords, at least in the
current implementation of the API\@. In addition, an attacker would not need both
individuals to be in the confused deputies group to deduce distance\footnote{In fact, in GAEN, each side is able to deduce distance independently, with the Bluetooth metadata merely providing calibration information ensuring better precision.}. The quality of the inference will decrease, but in contrast SwissCovid learns \emph{nothing} of contacts that occur with non-users. In other words, an attacker's view on close contacts is $\sim \alpha_{CD}$.
\item[infected individuals:] An attacker would obviously benefit from the
public exposure of the Temporary Exposure Keys: any SwissCovid user who is
infected would be known to them. Some may be in the confused deputies group while
others may have their beacons and MAC addresses broadcasted to someone in
the confused deputies group, which is then sent to the attacker.
Additionally, the attacker would have some
ability to figure out who in the confused deputies group is infected by
examining their location data for conspicuous patterns consistent with an
infection (e.g., go to a testing centre, then head home and stay there). An attacker would be blind to anyone they cannot track, but who has been identified by health authorities to be infected. Note that this disadvantage of the attacker would counterintuitively weaken as $\alpha_{SC}$ grows! 
\item[at-risk contacts:]
Because of both effects above, it seems likely that an attacker would quickly have better knowledge than
SwissCovid or Swiss health authorities, both in coverage and accuracy, of
at-risk contacts that took place. Indeed, we end up essentially comparing $\alpha_{SC}^2$ to $\alpha_{CD}$, modulo factors that are of comparable sizes in the two situations. 
\end{description}
We stress that this is a  coarse analysis of the situation, that should
definitely be refined. We observe, however, that this is concerning, as an
attacker with such control over the informational layer above the
epidemiological layer could start intervening in ways that preserve their
advantage. Part of the advantage might be in keeping authorities confused as to what is happening with the system, and creating tensions as a result of this confusion.

One might wonder what would be the harms in an attacker having deeper knowledge
of this kind. We \emph{believe} that this knowledge would translate into an economic
advantage to the attacker - if they so wish - by means of anticipating trends in economic disruption. Additionally, we
\emph{believe} that there is a large potential for disruption that can be achieved
from this position. This disruption might again be economically motivated, but
could also have geostrategic motives around disinformation and information
warfare. Historically, such mixed situations have proven very potent, as they
provide state-level actors a rich ecosystem of economically motivated
intermediaries to act (partly) on their behalf, or at least provide cover and
deniability. 

\section{Conclusion}
\textbf{We encourage the Swiss authorities} to re-evaluate the risks associated to the threats previously submitted to them, as we simply do not think their initial assessment is reflective of the real situation. We further encourage the Swiss authorities to communicate more clearly with each individual user on those risks, if they wish to remain coherent in their logic of providing information to Swiss citizens and residents so they are best able to make decisions in these difficult times. 

As they set parameters to their GAEN framework, \textbf{we encourage Apple and
Google} to keep in mind the threat of third party surveillance,
especially out of the older phones running on their OSes. We remind them
of the disproportionate power they have in detecting apps participating
in this ecosystem of Bluetooth surveillance, and their role in approving
them in the first place. A starting point is clearly to extensively
audit apps that require the relevant Bluetooth permissions:
a compass app that requires administrative privileges over the Bluetooth stack
does seem to violate the very essence of a least-privilege-based permission system.

\textbf{We encourage Apple, Google but also other big actors of the advertising ecosystem} such as the Internet Advertising Bureau or Facebook to constructively work with different national data protection authorities in documenting this ecosystem of Bluetooth surveillance. 

\textbf{We encourage the Swiss Data Protection Commissioner} to reevaluate the protocol underpinning the GAEN framework, and in particular to reconsider the RPI continuously emitted through the protocol as personal data, in light of Swiss jurisprudence cases \textsf{1C\_230/2011 ; ATF 138 II 346} (\emph{Google Inc. et 
Google Switzerland S.\`{a}.r.l. c.  Pr\'{e}pos\'{e} f\'{e}d\'{e}ral \`{a} la  protection des donn\'{e}es  et \`{a} la transparence}) on the publicity and identifiability of data published through \emph{Google Street View}.
\bibliographystyle{ieeetr}
\bibliography{references}
\end{document}